\documentclass[]{article}
\usepackage[letterpaper, margin=1in]{geometry}
\usepackage{fixltx2e,fix-cm}
\usepackage{epigrafica}
\usepackage[LGR,OT1]{fontenc}
\usepackage{amssymb}
\usepackage{bm}
\usepackage{graphics}
\usepackage{graphicx}
\usepackage{lscape}
\graphicspath{{./Figures/}}
\usepackage{mathptmx,txfonts}
\usepackage{theorem}
\usepackage{times}
\usepackage{makeidx}
\usepackage{pdfpages}
\usepackage{rotating}  

\newcommand{\ket}[1]{|#1\rangle}

\usepackage{color}  

\newcommand{\kr}[1]{|\mathrm{#1}\rangle}

\begin{document}

\Large{{\textbf{On the implications of the ``Quantum-Pigeonhole Effect''} \normalsize \\\

Alastair Rae and Ted Forgan

School of Physics and Astronomy, University of Birmingham, Edgbaston, Birmingham, B15 2TT, UK.\\\

\textbf{There has been considerable interest in a recent preprint \cite{php} describing an effect named as the ``Quantum Pigeonhole Principle''. The classical pigeonhole principle (classical PHP) refers to a result in number theory \cite{CPHP} which states that if $n$ objects are distributed between $m$ boxes, with $m < n$, then at least one box must contain more than one object. An experiment is proposed in \cite{php} where interactions between particles would reveal that they were in the same box, but a quantum mechanical measurement would imply that no more than $1$ of the $n$ objects is contained in any of the $m$ boxes, even though $n>m$. This result has been greeted by the authors of \cite{php} and some others as being of great importance in the understanding of quantum mechanics \cite{pworld,NScien}. In this paper we show by a full quantum mechanical treatment that the effect appears to arise as a result of interference between the components of the wavefunctions, each of which is subject to the classical PHP.}\\

 An experiment designed to demonstrate the Quantum Pigeonhole Principle -- herein referred to as the quantum pigeonhole effect or quantum PHE -- is proposed in Ref.~\cite{php} where three particles (such as electrons) pass simultaneously through two arms of an interferometer, which provide the ``pigeonholes". If either two or three pass together through the arms, they are expected to repel each other (e.g. by Coulomb repulsion), while particles in different arms are assumed to be too far apart to interact significantly.  The interaction within an arm will result in the particle paths being deflected and detecting this deflection should provide information about whether or not the particles have passed together along the arms.  Ref. \cite{php} predicts that if one ``post-selects'' those cases where all three particles are detected in the same output channel of the interferometer, the result implies that no interactions have occurred between any of the particles and, by implication, the classical PHP has failed.

In the present paper, we perform a detailed analysis of the case of three interacting particles passing through an interferometer with the aim of identifying which properties of this arrangement produce results that appear to imply the quantum PHE as well as those that do not.  We find that all results can be derived by calculations using quantum mechanics applied to states that represent the particles exploring all possible combinations of paths allowed by the classical PHP.  If the interactions are strong enough to deflect the particles appreciably, then the results directly confirm the classical PHP  \cite{CPHP}. However, if the interactions are sufficiently weak, then any deflection can only be observed by many repeated measurements and such ``weak measurements'' \cite{weak} may \emph{simulate} the quantum PHE result, even though they arise from states obeying the classical PHP. We also note that the interactions between the particles introduce phase shifts in their wavefunctions, which were not considered in \cite{php}, and we find that these have important consequences for the observability of the quantum PHE.
\section *{Results}

The apparatus for the proposed experiment is a Mach-Zehnder interferometer (MZI) shown in Fig.~\ref{fig1} in which each beam of quantum particles is split by one 50\% beam splitter (BS$_1$) and recombined by another (BS$_2$).  The two arms of the MZI are respectively labelled left (L) and right (R). If the apparatus is set up so that the L and R path lengths are equal, then all the particles will arrive at detector D$_A$. If however a phase shift of $\pi$ is introduced into one arm (e.g. by displacing one of the reflectors)  then all particles arrive instead at D$_B$. If, as proposed in Ref.~\cite{php}, a phase shift of $\pi/2$ is introduced into the R arm then particles will have equal probabilities of arriving at D$_A$ or D$_B$.

Quantum mechanically, after BS$_1$ the wavefunction of a particle can be represented as $(\kr{L} + \kr{R})$.  (For simplicity, we omit normalization factors throughout this paper.) BS$_2$ then forms a coherent superposition of the wavefunctions travelling down the L \& R arms, such that the amplitudes arriving at D$_A$ and D$_B$ are the sum and difference respectively of the $\kr{L}$ and $\kr{R}$ amplitudes. With no additional phase shift all particles arrive at D$_A$, while if a phase shift of $\pi/2$ is introduced in R, the the amplitudes arriving at D$_A$ and D$_B$ are equal in magnitude, in agreement with the discussion above.

\subsection*{The two-particle case}
Before discussing the three-particle situation described in Ref.~\cite{php}, it is instructive to  consider the case of two interacting particles passing simultaneously through the modified MZI. We assume that these are confined to beams that do not overlap, in which case the particles are distinguishable, even if they are identical, so their joint wavefunction may be expressed as a simple product. (The extent of the particle wavefunctions is discussed further below and in \cite{SI}.)  After the particles pass through BS$_1$, their wavefunction is:
 \begin{equation}
    \ket{\Psi(1,2)} = \Big(\kr{L(1)} + \kr{R(1)}\Big)\Big(\kr{L(2)} + \kr{R(2)}\Big) = \kr{L(1)} \kr{L(2)} + \kr{R(1)} \kr{R(2)} + \kr{L(1)} \kr{R(2)} + \kr{R(1)} \kr{L(2)}, \label{eq10}
  \end{equation}
where the arguments refer to the coordinates of the distinguishable particles labelled 1 \& 2.  After the phase shifter this becomes:
  \begin{equation}
    \ket{\phi(1,2)} = \kr{L(1)} \kr{L(2)}  - \kr{R(1)} \kr{R(2)} + i\Big(\kr{L(1)} \kr{R(2)} + \kr{R(1)} \kr{L(2)}\Big),\label{eq20}
   \end{equation}
where the operation of  the phase shifter multiplies each $\kr{R}$ component by $i$. The amplitude of this state function determines the probability of both particles being detected at D$_A$.  We note the cancellation of the first two terms, which correspond to the particles travelling down the same arm as each other.
This means that if both particles are detected at D$_A$, their wavefunction corresponds to them travelling down different arms from each other. This is also true if both particles are detected at D$_B$, where the component detected is $(|L> - i|R>)$. We emphasise that ``different arms" corresponds to a superposition of both possible arms for particle 1, combined in each case with the opposite arm for particle 2; without this superposition, there would be no interference at BS$_2$.

It may similarly be shown that if one particle is detected at each detector, then their wavefunction corresponds to a superposition of states where they travel \emph{together} down the arms of the MZI. These are remarkable results because two-particle correlations can be revealed merely by selecting where the particles are detected. However to confirm this experimentally, the particles must interact with each other to reveal if they share the same arm of the MZI. By ``interact" we mean by long range forces, with no overlap of wavefunctions.

 As shown in Fig.~\ref{fig2}, if particles travelling down the same arm repel each other, then, assuming the interactions in each arm are identical, the geometry of the MZI assures that the two paths for each particle still meet and interfere coherently at BS$_2$. However, the wavefunctions in equations~(\ref{eq10}) and (\ref{eq20}) must now be modified to take account of these interactions. We use a notation where, for example, the state of particle $1$ after it has interacted with particle $2$ in the L arm is represented by $\kr{L_2(1)}$, whereas the case where there is no interaction with another particle is represented by  $\kr{L_0(1)}$. Hence the amplitude arriving at detector D$_A$ becomes: 

\begin{equation}
\ket{\Phi^\prime(1,2)} = \kr{L_2(1)}\kr{L_1(2)}  -\kr{R_2(1)}\kr{R_1(2)} + i\Big(\kr{L_0(2)}\kr{R_0(2)} + \kr{R_0(1)}\kr{L_0(2)}\Big) \label{eq30}
  \end{equation}

 We see that despite the interactions, the cancellation of the first two terms is exact, so the clear prediction in this case is that if both particles arrive at the same detector, they will not be deviated from their non-interacting positions. Conversely, if they are detected at separate detectors, they will both have been deviated by the interactions due to passing together down the same arms. Of course, even without this post-selection or the $\pi/2$ phase shift, one can always tell if the two particles travelled down the same arms as each other, by seeing if their paths are deviated. The interesting result from the modified MZI is that by post-selection, we can observe correlations between particle paths, while still maintaining ignorance about the individual L and R paths-- which is essential if interference between them is to occur. This point was made strongly in \cite{php}.  An implication of this is that ``pigeonholing" means a correlation between particle trajectories. For instance, any combination of the first two terms on the right-hand side of equation~(\ref{eq30}) represents particles in the ``same'' pigeonhole, and any combination of the last two terms represents particles in separate pigeonholes.  Thus a ``pigeonhole" does not in general represent distinct physical locations for the particle paths, because the particles explore both arms of the MZI.

The conundrum presented by the modified MZI with three particles present may be simply expressed: if we post-select events where all three arrive at the same detector, does this mean that they all travelled by different paths? This is perfectly consistent with the classical PHP for two particles, but for three in an MZI having only two arms it appears not to be!

\subsection*{Quantum-Mechanical treatment of the three-particle case}

We now consider the case where three identical interacting particles travel through the Mach-Zehnder interferometer illustrated in Fig.~\ref{fig1}.  They are assumed to travel simultaneously~\cite{SI} and to exert forces on each other if they pass through the same arm (R or L) of the MZI.  The initial state is such that, after passing through the first beam splitter (BS$_1$), the wavefunction representing the state of the three particles has the form

\begin{eqnarray}
 \ket{\Psi(1,2,3)} &=& \Big(\kr{L(1)} + \kr{R(1)}\Big) \Big(\kr{L(2)} + \kr{R(2)}\Big) \Big(\kr{L(3)} + \kr {R(3)}\Big)\nonumber\\
  &=& \kr{L(1)}\kr{L(2)}\kr{L(3)}+ \kr{L(1)}\kr{L(2)}\kr{R(3)}+ \kr{L(1)}\kr{R(2)}\kr{L(3)}+ \kr{L(1)}\kr{R(2)}\kr{R(3)}\nonumber\\
  & &+ \kr{R(1)}\kr{L(2)}\kr{L(3)}+ \kr{R(1)}\kr{L(2)}\kr{R(3)}+ \kr{R(1)}\kr{R(2)}\kr{L(3)}+ \kr{R(1)}\kr{R(2)}\kr{R(3)}\nonumber\\
  \label {eq40}
\end {eqnarray}

\noindent We note that all eight components of this superposition obey the classical PHP because in every case at least two of the three particles have travelled along the same arm.

When the particles move through the same arm of the MZI they interact, and alter their directions of motion.  Extending the notation used in the two-particle case, $\kr{L_2(1)}$ represents the state of particle 1 after it has passed along the L arm with particle 2, but not particle 3, while $\kr{R_{23}(1)}$ represents the state of particle 1 after travelling along the R arm with both the other particles.  After passing through the phase-shifter, all $\kr{R}$ components are multiplied by $i$ and the state of the three particles is then represented by $\ket{\Phi}$ where

\begin{eqnarray}
 \ket{\Phi(1,2,3})&=& \kr{L_{23}(1)}\kr{L_{31}(2)}\kr{L_{12}(3)} + i\kr{L_2(1)}\kr{L_1(2)}\kr{R_0(3)}+ i\kr{L_3(1)}\kr{R_0(2)}\kr{L_1(3)}\nonumber\\
 & &- \kr{L_0(1)}\kr{R_3(2)}\kr{R_2(3)}
+ i\kr{R_0(1)}\kr{L_3(2)}\kr{L_2(3)}- \kr{R_3(1)}\kr{L_0(2)}\kr{R_1(3)}\nonumber\\
  & &- \kr{R_2(1)}\kr{R_1(2)}\kr{L_0(3)}-i\kr{R_{23}(1)}\kr{R_{31}(2)}\kr{R_{12}(3)}\nonumber\\
  & &\label {eq50}
\end {eqnarray}

 After passing through the second beam splitter, particles in corresponding $\kr{L}$and $\kr{R}$ states will constructively interfere to form states of the type $\ket{\phi} = \kr{L} + \kr{R}$ at detector D$_A$.  (The particles may continue to repel each other while they travel from BS$_2$ to the detector, but this is ignored from now on -- i.e. it is assumed that the distance from BS$_2$ to D$_A$ is small compared with the length of the path through the MZI, or that the particle interactions are screened in this region.) Using equation~(\ref{eq50}), the final state at D$_A$ is then:

\begin{eqnarray}
\ket{\Phi(1,2,3)}& = &\Big(1-i \Big)\times \Big(\ket{\phi_{23}(1)}\ket{\phi_{31}(2)}\ket{\phi_{12}(3)}-\ket{\phi_2(1)}\ket{\phi_1(2}\ket{\phi_0(3)}\nonumber\\
& & -\ket{\phi_3(1)}\ket{\phi_0(2)}\ket{\phi_1(3)}-\ket{\phi_0(1)}\ket{\phi_3(2)}\ket{\phi_2(3)}\Big)\label{eq60}
\end{eqnarray}

\noindent All the component products in equation~(\ref{eq60}) represent states where at least two particles are in either the right or left arm, which is consistent with the classical PHP.

 The only actual measurement made in this experimental setup is of the positions  of the particles when they are detected at D$_A$.  Figure~\ref{fig3} shows the expected arrival positions of the particles at D$_A$ in a plane perpendicular to the beam direction.  We have assumed that the configuration of the three particle beams when they passed through BS$_1$ formed an equilateral triangle.  We note the expected three-fold symmetry.

 We can now calculate the probability density for particle 1 being being detected in the vicinity of the position $\mathbf{r}_1$.  This is given by $P(\mathbf{r_1})d^2\mathbf{r}_1$ where

  \begin{eqnarray}
P(\mathbf{r}_1)& = &\int\int | \Phi(\mathbf{r}_1,\mathbf{r}_2,\mathbf{r}_3) | ^2d^2\mathbf{r}_2d^2\mathbf{r}_3 \Bigm/
 \int\int\int | \Phi(\mathbf{r}_1,\mathbf{r}_2,\mathbf{r}_3) | ^2d^2\mathbf{r}_1d^2\mathbf{r}_2d^2\mathbf{r}_3  \nonumber\\
 & = &|\phi^{}_{23}|^2 + |\phi^{}_{2}|^2 + |\phi^{}_{3}|^2 + |\phi^{}_{0}|^2\nonumber\\
 & & -\Big( \phi^{*}_{23}\phi^{}_{2}S_{31,1}S_{12,0} +\phi^{*}_{23}\phi^{}_{3}S_{31,0}S_{12,1} +\phi^{*}_{23}\phi^{}_{0} S_{31,3}S_{12,2} + \mathrm{ c.c. }\Big) \nonumber\\
 & & +\Big( \phi^{*}_{2}\phi^{}_{3} S_{1,0}S_{0,1}  + \phi^{*}_{2}\phi^{}_{0} S_{1,3}S_{0,2} + \phi^{*}_{0}\phi^{}_{3} S_{3,0}S_{2,1}  + \mathrm{ c.c. } \Big)  \label{eq70}
\end{eqnarray}

 \noindent We are now using a position representation where, for example, $\phi_0 = \phi_0(\mathbf{r}_1)$, $\phi_2 = \phi_2(\mathbf{r}_1)$ and $\phi_{23} = \phi_{23}(\mathbf{r}_1)$.  The quantities represented by $S$ are overlap integrals with the suffixes indicating which types of $\phi$ are involved: e.g. $S_{31,1}=\int\phi^{*}_{31}(\mathbf{r})\phi^{}_1(\mathbf{r})d^2\mathbf{r}$ and  $S_{0,1}=\int\phi^{*}_0(\mathbf{r})\phi^{}_1(\mathbf{r})d^2\mathbf{r}$.  Details of the calculations of the cross terms are given in Ref.~\cite{SI}. However, we note that equation~(\ref{eq70}) has a simple form both in the limit where the interaction is zero and all the $S$s are unity, and in the opposite strong interaction limit where the deflection is sufficiently large that there is no overlap, and the $S$s are zero. In both cases, the sum of the last six terms is zero and the probability density is just the incoherent sum of the contributions from the four $\phi$s.


\subsection*{Numerical calculations}

Using equation~(\ref{eq70}) with the triangular geometry described above, we have computed $P(\mathbf{r}_1)$ for a number of different interaction strengths. We have taken the $\phi$s to be Gaussians with the same widths, and assumed for the present, as in~\cite{php}, that the interactions do not lead to phase changes in the wavefunctions.
We define the interaction strength ($d$) as the magnitude of the expected deflection of one particle by another (see Fig.~\ref{fig3}) in units of the standard deviation of $|\phi|^2$. Typical results are shown in Fig.~\ref{fig4}.  We note that when the interaction is strong, the probability density separates into four distinct peaks corresponding to the different experiences undergone by the particles and the cross terms in the probability density have relatively little effect. This means that the detection of the position of, say, particle 1 would unambiguously determine whether it had been accompanied by particle 2, particle 3, both 2 and 3, or neither in its passage through the MSI.  This result is completely consistent with the classical PHP.

 In the opposite limit of no interaction, there is only one peak for each particle and it is of course centred on the original particle position.  This measurement clearly provides no information as to how many particles have populated the arms of the MZI.  That is, there is no information one way or the other about the PHP, and certainly no evidence that the classical PHP fails.

 The most interesting situation is where the interaction is weak and we illustrate this in Fig.~\ref{fig4} by the case where $d=0.25$.  Little if any information could be deduced from the detection of a single particle in this case, but in principle at least, this probability distribution could be measured by detecting the arrival positions of a large number of identically prepared triplets of particles.  As this probability distribution is very similar to that seen in the zero-interaction case, an experimenter might be tempted to conclude that the particles had not interacted and therefore must have all passed through the MZI unaccompanied -- which is the quantum PHE prediction in Ref.~\cite{php}.  However, if the results were interpreted with knowledge of the quantum analysis set out above, it would be realised that such a conclusion would be invalid because this feature of the observed particle density is largely or entirely a result of interference between various components of the wavefunction, all of which are subject to the classical PHP.    We investigated this point further by calculating the expectation value, $<y>$, of the displacement in the $y$ direction (see Fig.~\ref{fig3}), as a function of interaction strength.  (This could be obtained experimentally from the particle density recorded in a series of experiments with different interaction strengths.)   The results are shown in Fig.~\ref{fig5} along with the ``incoherent'' result that would be expected if there were no interference - i.e. ignoring all but the first four terms on the right-hand side of equation~(\ref{eq70}).  The behaviour in the latter case is a straight line whose slope is a measure of the displacement in the $y$ direction of the average position of the centres of the first four terms in equation~(\ref{eq60}).  What would actually be observed, however, is the expectation value calculated using the full wavefunction, $\Phi$.  We see that in this case the linear slope is zero at $d=0$, so such a result could lead to the conclusion that the particles had not interacted. Once again, however, an understanding of the full quantum calculation would vitiate this conclusion.  Information about the interaction strength could still extracted from this data, but this would involve an analysis of the quadratic (and possibly higher-order) dependence of $<y>$ on $d$. Another way of interpreting these results is that they correspond to a ``weak measurement'' of the slope of $<y>$ vs $d$, but it would clearly be wrong to equate this to a weak measurement of the interactions.

  It follows, therefore, that all these results can be derived using the classical PHP and they \emph{simulate}  the quantum PHE at small $d$ only because  there is interference between certain terms in the wavefunction. The ingenious arrangement in Ref. \cite{php} ensures that this occurs in the limit of weak interactions; however, destructive interference between the amplitudes of different trajectories does \emph{not} imply that these trajectories were forbidden. As noted in \cite{weak}, ``weak" measurements do not necessarily represent physical variables, and although they obey their own logic, they can lead to apparently paradoxical results. We can imagine an experimenter obtaining a small-$d$ result being drawn to conclude that this confirms the quantum PHE.  If however, she were aware of the results of the calculations in this paper, she would realise that the results arise from a superposition of quantum states, all of which are subject to the classical PHP.  She could indeed go further by fitting her results to the calculated curves, so deriving the actual interactions experienced by the particles. The quantum PHE therefore adds little new to our understanding of quantum mechanics, nor of number theory.

 It is also instructive to consider the case, briefly mentioned in \cite{php}, in which one post-selects the case where one particle arrives at one detector and two arrive at the other: for instance where particle 1 is detected at D$_A$ and particles 2 and 3 are detected at D$_B$. The final wavefunction in this case is:

\begin{eqnarray}
\ket{\Phi(1,2,3)}& = &\Big(1-i \Big)\times\Big(\ket{\phi_{23}(1)}\ket{\phi_{31}(2)}\ket{\phi_{12}(3)}+\ket{\phi_2(1)}\ket{\phi_1(2}\ket{\phi_0(3)}\nonumber\\
& & -\ket{\phi_3(1)}\ket{\phi_0(2)}\ket{\phi_1(3)}+\ket{\phi_0(1)}\ket{\phi_3(2)}\ket{\phi_2(3)}\Big)\label{eq80}
\end{eqnarray}

Figure~\ref{fig5} includes the results of our calculations for all three particles in this case . Unlike the situation where all three particles arrive at one detector, there is a linear dependence of $<y>$ on $d$ at small $d$ for particle 1. In fact, the slope near the origin of this graph is twice the incoherent result to which it tends at large $d$.  This initial slope corresponds to the magnitude of the deflection one would get if particle 1 were repelled on average by both particles 2 and 3.  If the only observation made were that of particle 1 arriving at  D$_A$ and if the interactions were weak, the same logic that led to the quantum PHP in the 3-particle case could lead an experimenter to believe that all 3 particles must have travelled together along the same arms of the MZI.  If, however, the experimenter also looked at D$_B$, she would observe the behaviour of particles 2 and 3  where there there is now a linear change in both $<x>$ and $<y>$ at small $d$.   The change in $<y>$ is the same for both particles and equals the incoherent value for all values of $d$.  The vector sum of all the particle displacements for all three particles is zero, so the three particles as a whole conserve transverse momentum. At large $d$,  $<x>$ for both  particles tends to zero, in line with the incoherent behaviour.  
This leads to the conclusion that the displacements of all particles at smaller $d$ are the result of interference between the different terms in equation~(\ref{eq80}).

 We now consider the results at small $d$ from the quantum PHE point of view. Since particles 1 and 2 are detected at \emph{different} detectors, our earlier discussion of the two-particle case leads us to expect that they travelled in the \emph{same} arms as each other. The same applies to particles 1 and 3, hence at small $d$ the state of particle 1 is formed by constructive interference of the two states corresponding to repulsion by each particle, which is what our calculation indicates; at large $d$, the two states do not overlap, constructive interference does not occur and the behaviour is the same as in the incoherent case.  This is similar to the earlier case of all three particles detected in D$_A$, except in that case the interference was \emph{destructive} resulting in the zero slope at small $d$ discussed above. Moreover, since particles 2 and 3 arrive at the \emph{same} detector, our two-particle discussion predicts that they would have travelled by \emph{different} paths, so do not repel each other: they are only repelled by particle 1, which again agrees with our calculation.  Classically, particles 2 and 3 cannot both be in the same arm as particle 1, at the same time as being in different arms from each other, but quantum-mechanically they can be in a superposition of these states, which is the basis of our calculation.

\subsection*{The effects of the phase shifts in wavefunctions due to interactions between particles}

We now turn to another matter, not considered by the authors of [1], which greatly limits the possible practical observation of the quantum PHE. This is the question of phase differences between the components of the superposition representing the particle as it reaches the detector.  These result from two causes: the angular separation of the beams and the effect of the interaction potential.

A quantity relevant to both of the above is the width of the probability distribution $|\phi|^2$ for an isolated particle at the detector. This is determined by the width of the incoming beam and the spread due to diffraction over the distance $\ell$ travelled inside the MZI. Minimising the sum of these, it may be shown that the smallest possible beam width at the detector is given by:

 \begin{equation}
 \sigma = (\ell \lambda/2 \pi)^{1/2}, \label{F1}
 \end{equation}

\noindent where the variation of the probability density with radius $r$ from its centre is given by  $|\phi|^2 \propto exp(-r^2/2 \sigma^2)$ and $\lambda$ is the wavelength of the particle in the beam direction.

The dimensionless measure ($d$) of the strength of the interaction between particles as defined above can be expressed in terms of the expected deflection, $\Delta r$, of one particle by another (see Fig.~\ref{fig3}) and the standard deviation ($\sigma$) of $\phi^2$ as:

\begin{equation}
d = \Delta r / \sigma . \label{F2}
 \end{equation}

If two particles repel each other over the whole distance $\ell$, then each deviates parabolically from a straight line path, and arrives at the detector deflected by an angle $\alpha$  given by:

 \begin{equation}
 \alpha= 2 \Delta r /\ell. \label{F3}
 \end{equation}

The angle of incidence $\alpha$ leads to a variation of phase $\Delta \theta$ across the wavefunction $\phi$. Taking $s$ as the distance from the undeflected beam along the direction of deflection, we have:

 \begin{equation}
 \Delta \theta= 2 \pi (s  \alpha /\lambda ) =  4 \pi (s   \Delta r )/(\ell \lambda). \label{F4}
 \end{equation}

Using equations~(\ref{F1}) to (\ref{F3}), this may be written as:

 \begin{equation}
\Delta \theta = 2 (s / \sigma)   \times   (\Delta r / \sigma) = 2 (s / \sigma)   d. \label{F5}
 \end{equation}

We see that this phase difference is significant when the particle deflection is comparable with the spread of the wavefunction.

Turning now to the phase change due to the interactions, we denote this as $\Delta \theta_i$.   The total energy of a particle passing through the MZI is constant, which means that particles moving together along the same arms must do so at a lower speed than is the case when they move along different arms.  The effect of changes of potential energy on phase is well known from interferometry with neutrons in a MZI \cite{ndiff}.    We first consider two particles 1 and 2 travelling through the MZI.  If the particles travel together down the \emph{same} arm (it does not matter which one), they have a mutual potential energy that we shall denote by $\Delta V$ per particle. If they travel by different arms, the mutual interaction is by assumption negligible. Hence $\Delta V$ will give rise to a phase difference between the two components of the wavefunction of particle 1: those where it is accompanied by 2, and those where it is not. (The same will happen to the wavefunctions of particle 2, but the consequences for this particle are identical, so we do not consider it further).

For a non-relativistic particle of mass $m$, the magnitude of momentum change due to  $\Delta V$ is given by:

\begin{equation}
 \Delta V  = p \Delta p / m . \label{F6}
 \end{equation}

\noindent Hence, the phase difference introduced is:

\begin{equation}
  \Delta \theta_i = \ell \Delta p / \hbar = (\Delta V t/ \hbar ) .  \label{F7}
 \end{equation}

\noindent where  $t$ is the time the particles spend at the potential $\Delta V$.  $\Delta \theta_i$ is the phase difference between, e.g., $\phi_2(1)$ and $\phi_0(1)$.  (The phase difference between $\phi_{23}(1)$ and $\phi_0(1)$ would be double this.)

The displacement of a particle position due to interaction with another depends not on the total potential energy $2 \Delta V$ but on its gradient, so to make progress we assume a Coulomb interaction, which will be long-range enough to cause a repulsion between equally-charged particles separated by a distance $r$ that is considerably larger than their wavefunction spread $\sigma$. (This limitation is necessary so that the particles can to a good approximation be regarded as being deflected classically by a potential which is a function of the separation of their centres of mass.) In this case, the force between the particles has the magnitude $F = 2 \Delta V/r$. After travelling for a time $t$, the particle tracks are deviated by $\Delta r = \sigma d = \frac{1}{2} (F / m) t^2 =  (\Delta V/mr) t^2$. This may be rewritten as:

\begin{equation}
 d =  \Delta V t \ell /(mvr\sigma) =  \Delta V t \ell \lambda / (2\pi \hbar r\sigma) \label{F8}
 \end{equation}
		
\noindent Using equation~(\ref{F1}) for the beam width $\sigma$ and equation~(\ref{F7}) for the phase difference, we obtain:
	
\begin{equation}
 \Delta \theta_i  = r\Delta r / \sigma^2 =  (r/\sigma) \times d.              \label{F9}
 \end{equation}


We note that to avoid overlap between the wavefunctions of different particles, we need $r >> \sigma$, which implies that $ \Delta\theta_i >  2\pi$, unless $d$ is very small indeed. Hence for a Coulomb interaction, the phase difference introduced by the interaction cannot be ignored, and should be included in calculations to ascertain if the quantum PHE is observable. One could devise mathematical potentials that avoid or minimise the effects discussed, but these are unlikely to be physically realisable. The three particle case is very similar, but the potential energy difference for one particle passing through one arm with two in the other, and all three in the same arm is 2$\times\Delta V$ as defined above.

We have repeated our calculations for the case of all three particles arriving at one detector, with the phase change due to beam divergence given by equation~(\ref{F5}) while that due to the interactions (equation~(\ref{F9})) is taken as $5d$.  (The spacing of separate particle beams is then just $5 \times$ the spread of each beam.) The results are shown in Fig.~\ref{fig6} which shows plots of $<y>$ versus $d$ and plots of the probability distributions for two values of $d$. We note that oscillations have appeared in the dependence of $<y>$ on $d$ and the region where the apparent quantum PHE is observed is now confined to the region where $d\lesssim 0.005$, compared with $0.5$ when the phases are ignored (Fig.~\ref{fig4}).  This is consistent with the prediction made above.  The oscillations result from the inclusion of the interaction phase ($\theta_i$) in the calculation and the fact that the apparent quantum PHE effect depends on interference between the amplitudes of the different components of the wavefunction contributing to the cross terms in equation~(\ref{eq70}).  These are so large that $<y>$ is actually a little \emph{below} zero for $d\sim 0.25$ so the average of many measurements in an experiment performed with exactly these parameters at this value of $d$, could be misinterpreted as indicating that the particles were attracting each other! The oscillations die out for values of $d\gtrsim 0.7$.  This is due to the varying phase across the beams leading to increasingly variable phase differences between the wavefunction components. We note, in contrast to this 3-particle case, that our prediction for the 2-particle result does not vary with the strength of the interaction. The 2-particle case does not of course give rise to any pigeonhole paradoxes.

We have also investigated the possibility of carrying out the measurements described above using electrons.  This has been revealed to be extremely challenging by a calculation in \cite{SI} which shows that all measurements would have to be made on subatomic scales. Indeed, unless the phase differences are tightly controlled (and the finite size of the particle wave packets may affect this~\cite{SI}) the coherence between the wavefunctions representing the possible paths of each particle might well not be maintained so the probability density would have no cross terms in it and all traces of the quantum PHE would disappear.

\section*{Conclusions}
We have demonstrated that the passage of independent interacting particles through a Mach-Zehnder Interferometer with added $\pi/2$ phase shift may be derived from a correct application of quantum mechanics to the classical PHP, in which the particles explore all combinations of paths allowed by classical number theory. We show that all the apparently paradoxical results of the supposed quantum pigeonhole principle can be accounted for by our treatment, and that they only apply when the interaction between particles is extremely weak -- especially if the effects of phase shifts introduced by the inter-particle interactions are allowed for. In the weak interaction limit, results can only be obtained by a statistical analysis of many experiments, and the results of a single experiment are indistinguishable from the zero interaction case, while with stronger interactions, all results are consistent with the classical PHP. Everyone accepts that quantum mechanics is weird, but it is not as weird as the quantum pigeonhole principle!  Rather than being a new “Principle”, the quantum PHE is instead an elegant demonstration of the ineluctable non-locality of quantum mechanics.

\section*{Acknowledgements} We thank Elizabeth Blackburn and Igor Lerner for useful comments. E.M.F. acknowledges support from EPSRC Grant \#  EP/J016977/1.

\section*{Author contributions}
A.I.M.R. and E.M.F. both contributed to the calculations and physical interpretations in this paper.

\section*{Additional information}
Competing financial interests: The authors declare no competing financial interests.

\clearpage

\begin{figure}[p]
\begin{center}
\includegraphics[width=0.8\textwidth]{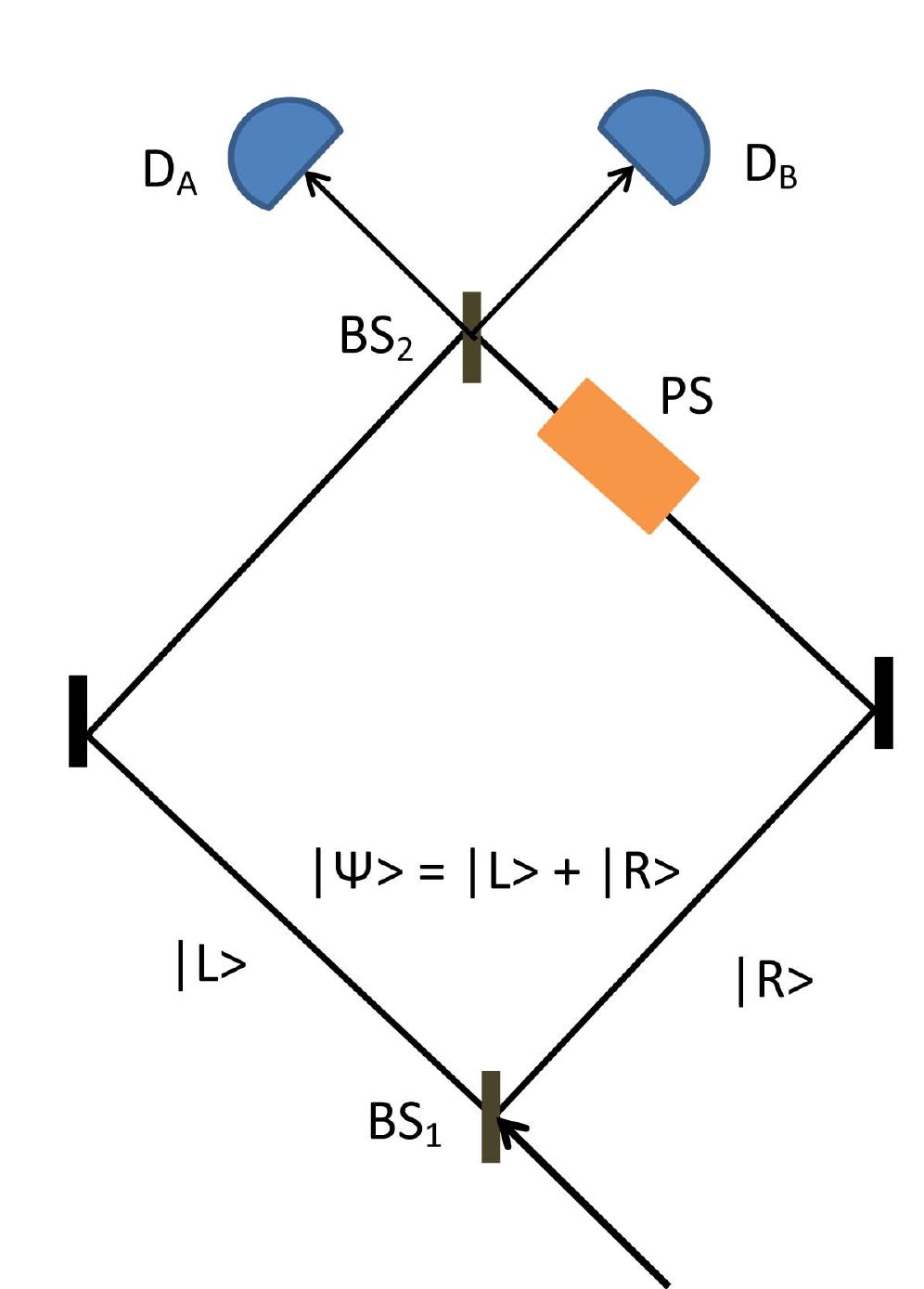}
\end{center}
\caption{\textbf{The Mach-Zehnder interferometer.}  BS$_1$ and BS$_2$ are 50\% beam splitters, D$_A$ and D$_B$ are detectors and PS is a phase shifter. $\ket{\psi}$ represents the state of a single particle emerging from BS$_1$. If the phase shift is zero, the particle is certain to arrive at D$_A$; if it is $\pi$, the particle is certain to arrive at D$_B$, while if it is $\pi/2$, the probabilities of detection at D$_A$ and D$_B$ are equal.}
\label{fig1}
\end{figure}

\begin{figure}[p]
\begin{center}
\includegraphics[width=0.8\textwidth]{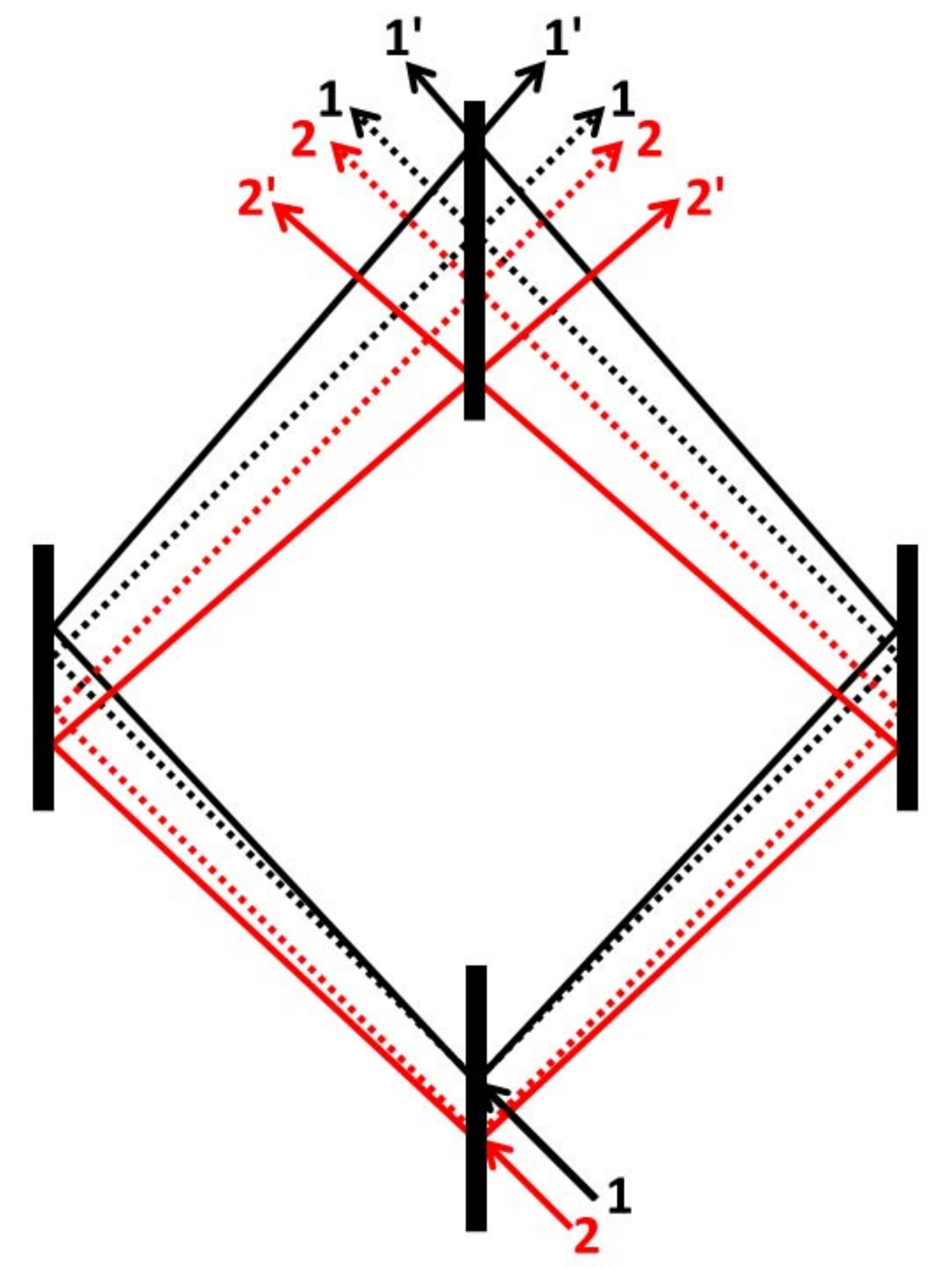}
\end{center}
\caption{\textbf{Schematic representation of the paths of two particles through a MZI}.  If the particles do not interact  (dashed paths, with un-primed labels), they arrive at the second beam splitter and the L \& R paths for each particle interfere in the usual way. If they interact, their paths diverge, so that they arrive at BS$_2$ with positions and angles displaced and changed path lengths  (solid lines, with primed labels). However, the symmetry of the MZI ensures that the L \& R paths of each displaced particle are still coherent with each other and can interfere.}
\label{fig2}
\end{figure}

\begin{figure}[p]
\begin{center}
\includegraphics[width=0.8\textwidth]{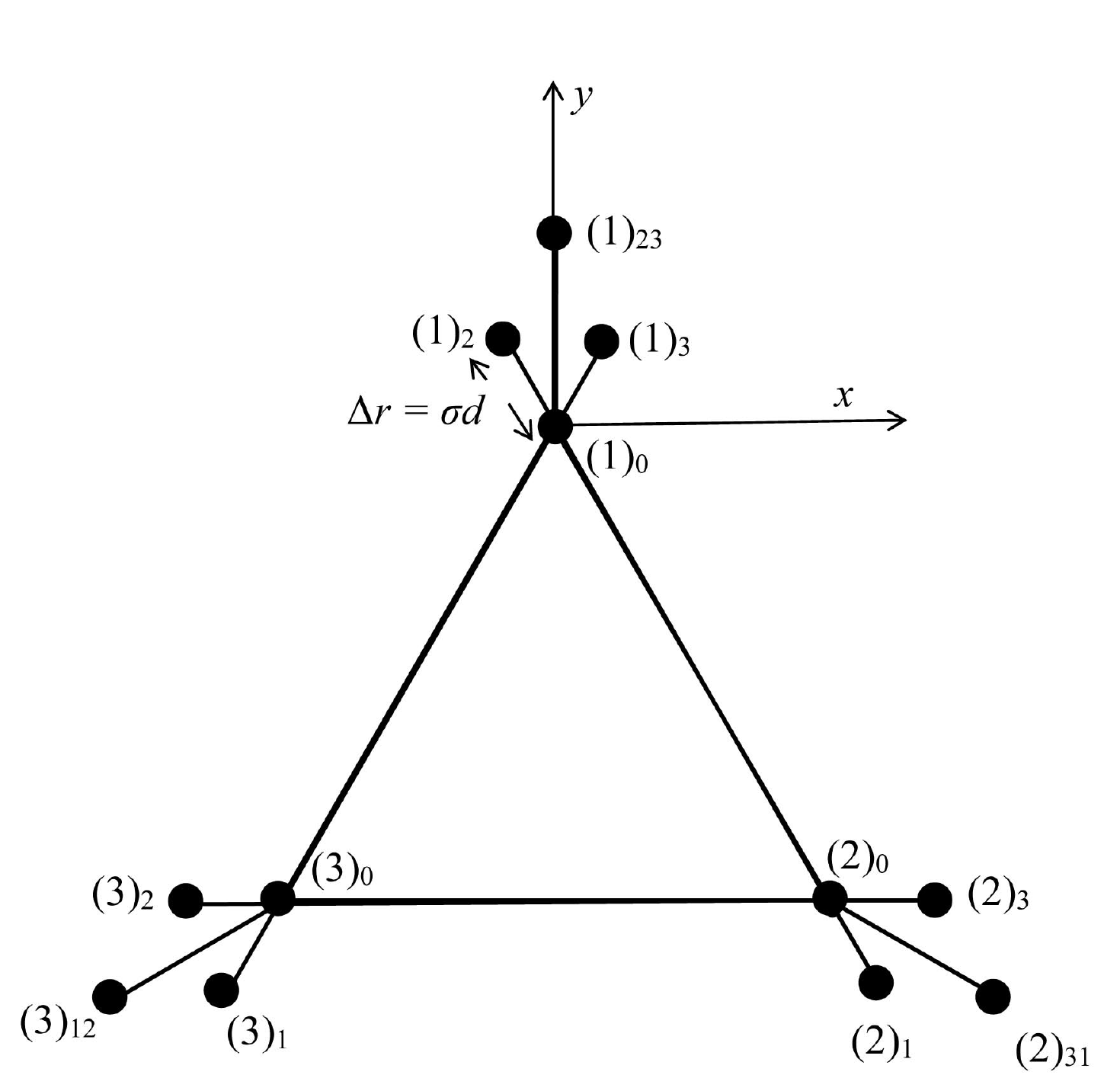}
\end{center}
\caption{\textbf{Three particles in the MZI} The relative positions of the three beams are represented by the filled circles at the vertices of an equilateral triangle.  After interacting repulsively, they are expected to be displaced into the positions shown.  The notation is similar to that used in the text:  e.g. $(1)_2$ represents the position of particle 1 after it has passed through the MZI in company with 2, but not 3, while $(1)_{23}$ represents the position of 1 after being accompanied by both both 2 and 3, and $(1)_0$ denotes an undeviated particle. The distance $\Delta r = \sigma d$ is proportional to the interaction strength.  The axes $x$ and $y$  are used in the later discussion (for particles 2 and 3, their $y$ co-ordinates similarly point away from the centre of the triangle).}
\label{fig3}
\end{figure}

\begin{figure}[p]
\begin{center}
\includegraphics[width=0.8\textwidth]{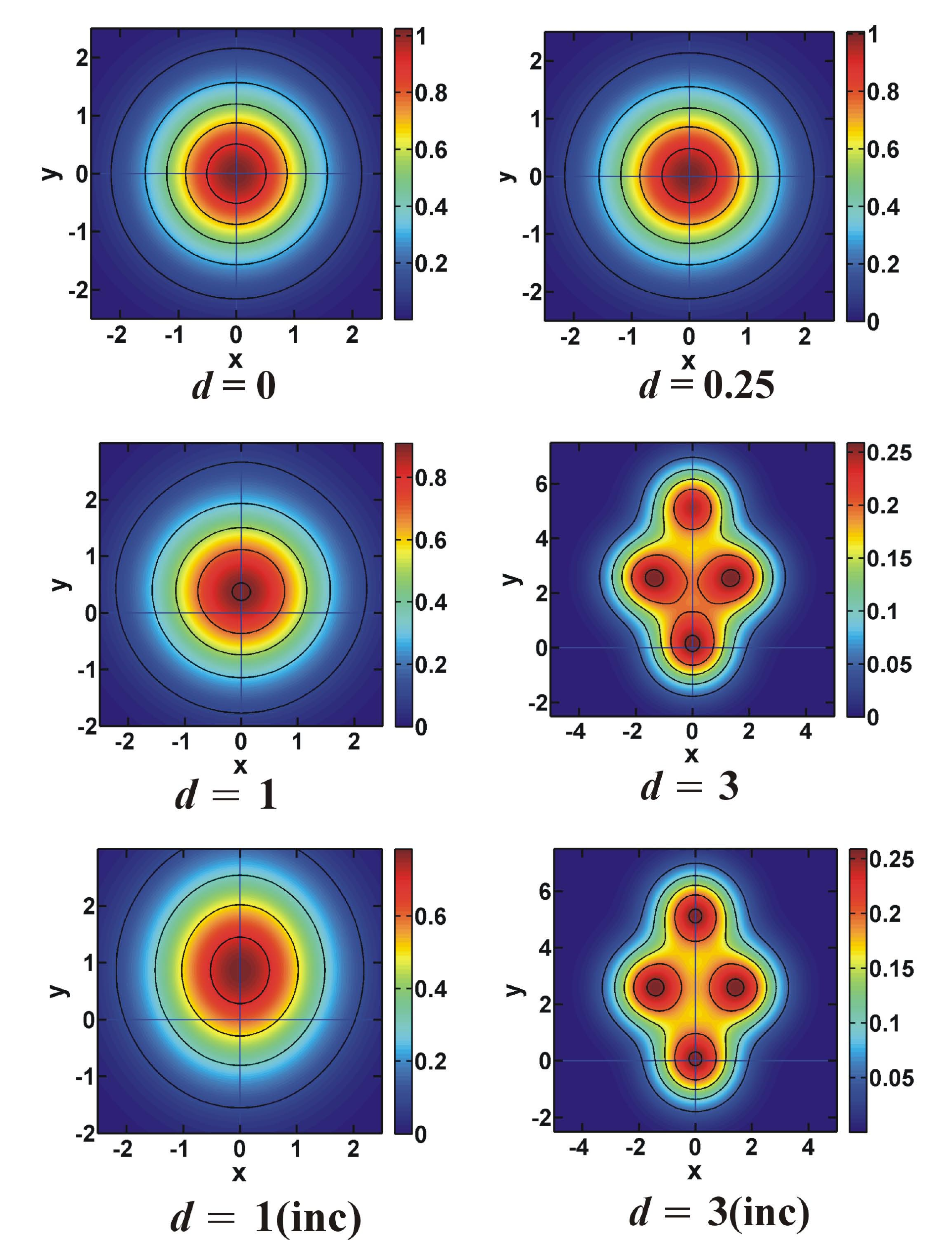}
\end{center}
\caption{\textbf{Probability densities for detection of particle 1} The top four panels represent calculations for the case when all three particles are detected at detector D$_A$, for various values of the interaction strength, represented by $d$ = $\Delta r /\sigma$. The last two panels (marked "inc") show for comparison the results if one ignores cross terms in the probability density, for two values of $d$; this is what would be observed if the contributions from different trajectories were incoherent. In all cases, the $x$ \& $y$ scales are in units of $\sigma$ and the density scale of the plots is set so that the peak value equals 1 for $d=0$ .  The plot areas for $d=3$ cover twice the range of $x$ \& $y$ used in the other plots.}
\label{fig4}
\end{figure}

\begin{figure}[p]
\begin{center}
\includegraphics[width=0.6\textwidth]{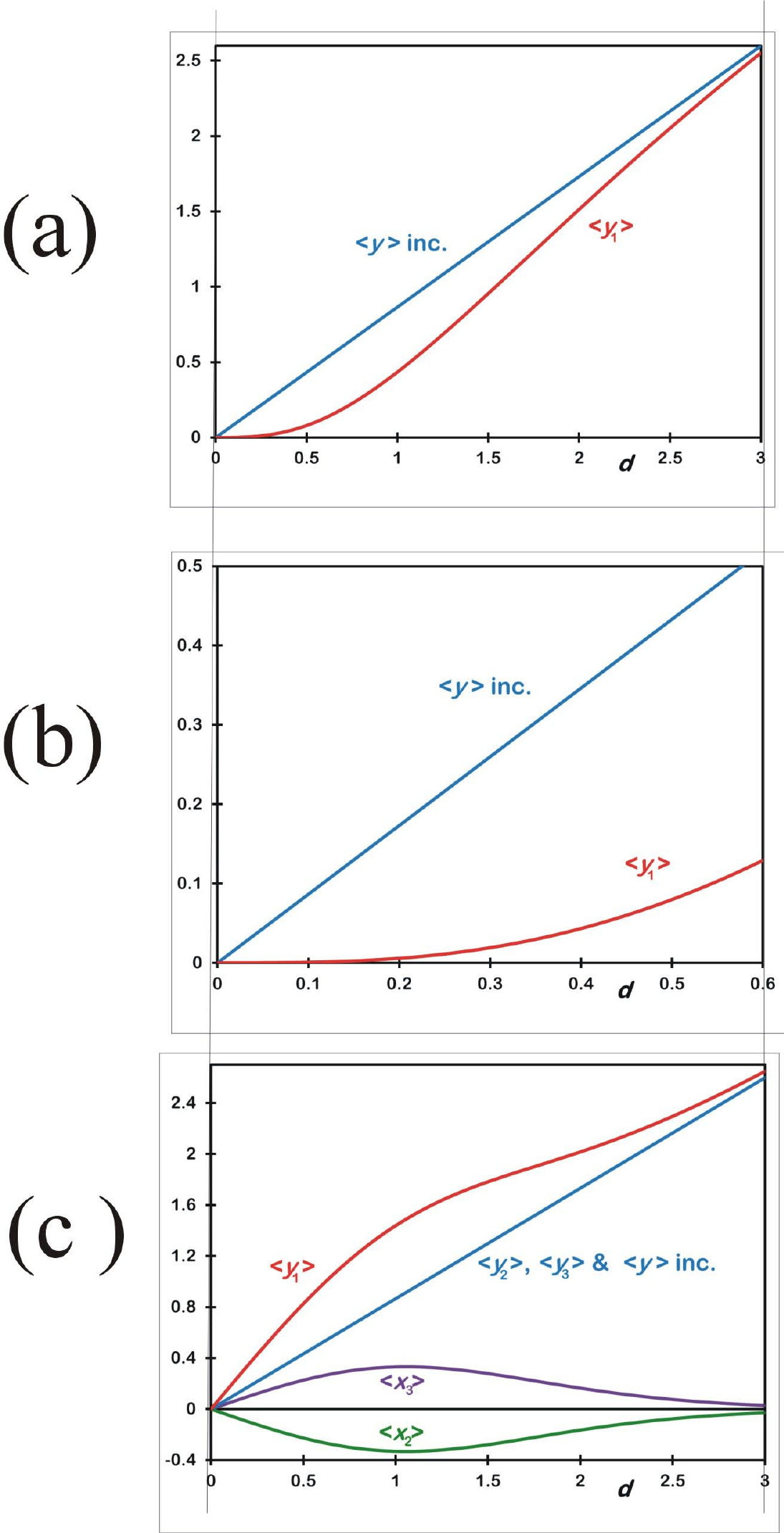}
\end{center}
\caption{\textbf{The  expectation values of the particle displacements} (a) shows $\langle y_1 \rangle$ versus interaction strength $d$ (both in units of $\sigma$), for events where all three particles arrive at the same detector. In this case, particles 2 \& 3 have the same $\langle y \rangle$ as particle 1, and all have zero $\langle x \rangle$. The numbered subscripts refer to particle labels and the label ``inc.'' denotes the  ``incoherent" result.  (b) shows the small-$d$ region of (a). (c) shows the equivalent calculations for the case when particle 1 arrives at one detector and the other two particles arrive at the other.}
\label{fig5}
\end{figure}

\begin{figure}[p]
\begin{center}
\includegraphics[width=0.8\textwidth]{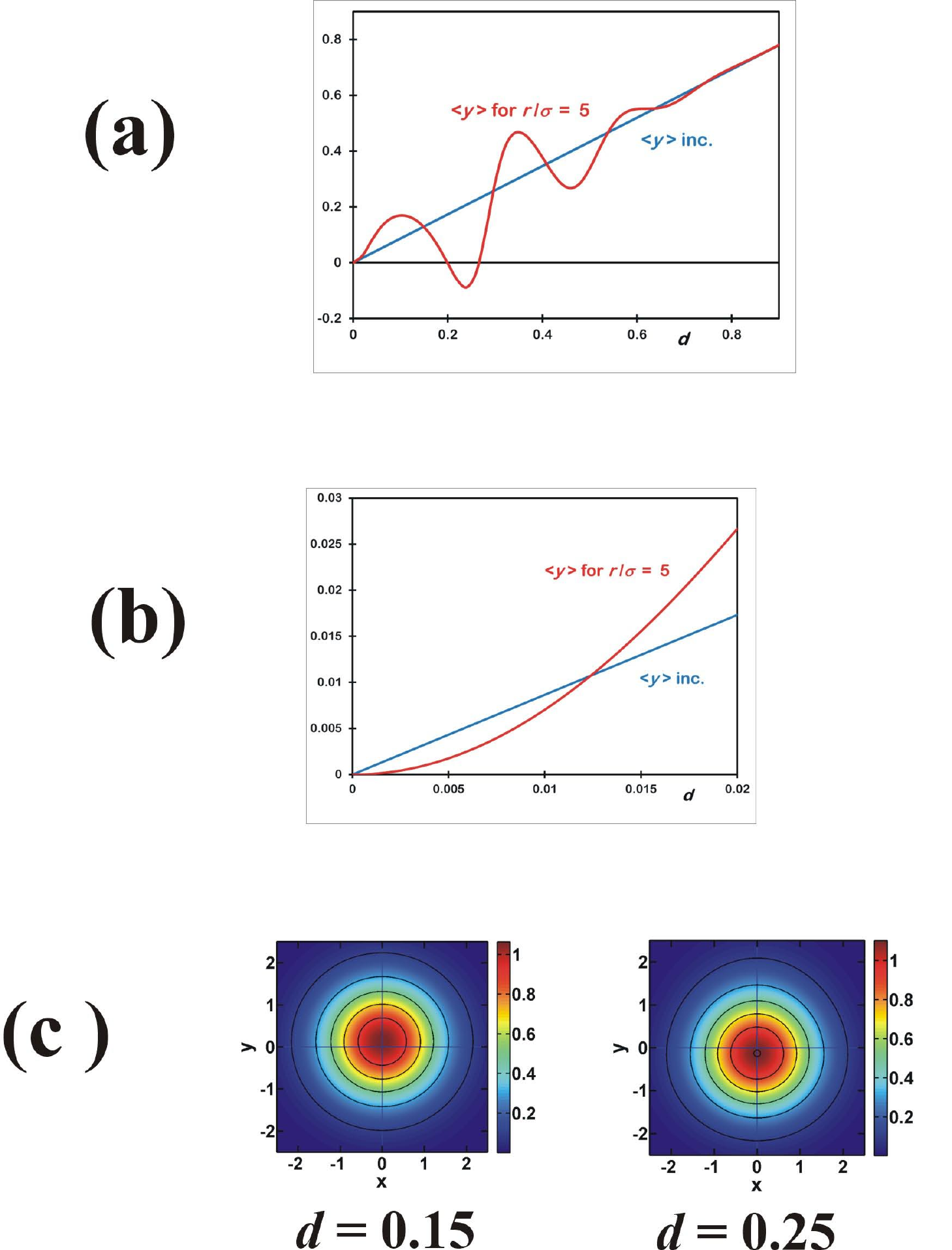}
\end{center}
\caption{\textbf{ Expectation value of the particle displacements with phase shifts included} (a) shows calculated $\langle y \rangle$ (in units of $\sigma$) versus interaction strength $d$, calculated using the expression~(\ref{F5}), with a value of interaction phase shift $\Delta\theta_i = r/\sigma \times d =5d$, for the case where all three particles are detected in D$_A$, so all have the same displacements.  The results for larger values of $ r/\sigma$ are similar except that the ``wavelength'' of the oscillations is reduced.  (b) shows the small-$d$ region of (a). We note that the region where the quantum PHP could be observed is now confined to extremely small values of $d$. (c) shows the probability density for two values of $d$.  The centre of the probability distribution varies non-monotonically with $d$, in agreement with the corresponding values of $\langle y \rangle$ shown in (a). The probability distribution for $d=3$ (not shown) is indistinguishable from the ``incoherent" result shown in the last panel of Fig.~\ref{fig4}. }
\label{fig6}
\end{figure}

\clearpage

\section*{Supplementary Information}

\subsection*{Extent of particle wavefunctions}

  We emphasised in the main text that if the particles are identical, it is necessary that there be no overlap between their wavefunctions so as to avoid complications associated with indistinguishability. However, the absence of overlap is also necessary so that the particles can to a good approximation be regarded as being deflected classically by a potential which is a function of the separation of their centres of mass.
This again requires that $r$, the distance between the parallel beams of particles must satisfy $r >>\sigma$, where $\sigma = (\ell \lambda / 2 \pi)^{1/2}$ -- see equation~(\ref{F1}) of the main text. If the beams are on diverging paths, $\ell $ is approximately the distance over which the particles are about $r$ apart. If the beams are focused onto the detector, then $\ell$ approximately equals the distance the particles have travelled before the focusing element.

In addition, the particles have to pass \emph{together} through the MZI, entering at the same time and with wavepackets of short length, comparable with $\sigma$. This implies that the wavelengths of their Fourier components must be spread over a range where $\delta\lambda/\lambda \geq \lambda/\sigma =  (2\pi \lambda / \ell )^{1/2}$. However, since this range can be small, it is no impediment to the operation of the MZI  if it is operated close to exact symmetry of L and R arms.

\subsection*{Calculation of overlaps and probability distribution for particle 1}

Cross terms in $\Phi^{*}(\mathbf{r}_1,\mathbf{r}_2,\mathbf{r}_3) \times \Phi(\mathbf{r}_1,\mathbf{r}_2,\mathbf{r}_3)$ are

\begin{eqnarray}
-\Big( \phi^{*}_{23}(1)\phi^{}_{2}(1) \times \phi^{*}_{31}(2) \phi^{}_{1}(2) \times \phi^{*}_{12}(3) \phi^{}_{0}(3) + \mathrm{ c.c. } \Big) \nonumber\\
-\Big( \phi^{*}_{23}(1)\phi^{}_{3}(1) \times \phi^{*}_{31}(2) \phi^{}_{0}(2) \times \phi^{*}_{12}(3) \phi^{}_{1}(3) + \mathrm{ c.c. } \Big) \nonumber\\
-\Big( \phi^{*}_{23}(1)\phi^{}_{0}(1) \times \phi^{*}_{31}(2) \phi^{}_{3}(2) \times \phi^{*}_{12}(3) \phi^{}_{2}(3) + \mathrm{ c.c. } \Big) \nonumber\\
+\Big( \phi^{*}_{2}(1)\phi^{}_{3}(1) \times \phi^{*}_{1}(2) \phi^{}_{0}(2) \times \phi^{*}_{0}(3) \phi^{}_{1}(3) + \mathrm{ c.c. } \Big) \nonumber\\
+\Big( \phi^{*}_{2}(1)\phi^{}_{0}(1) \times \phi^{*}_{1}(2) \phi^{}_{3}(2) \times \phi^{*}_{0}(3) \phi^{}_{2}(3) + \mathrm{ c.c. } \Big) \nonumber\\
+\Big( \phi^{*}_0(1)\phi^{}_{3}(1) \times \phi^{*}_{3}(2) \phi^{}_{0}(2) \times \phi^{*}_{2}(3) \phi^{}_{1}(3) + \mathrm{ c.c. } \Big)
\end{eqnarray}

Leading to:

\begin{eqnarray}
P(\mathbf{r}_1)& = &\int\int | \Phi(\mathbf{r}_1,\mathbf{r}_2,\mathbf{r}_3) | ^2d^2\mathbf{r}_2d^2\mathbf{r}_3/
\int\int\int | \Phi(\mathbf{r}_1,\mathbf{r}_2,\mathbf{r}_3) | ^2d^2\mathbf{r}_1d^2\mathbf{r}_2d^2\mathbf{r}_3 \nonumber\\
& = &|\phi^{}_{23}|^2 + |\phi^{}_{2}|^2 + |\phi^{}_{3}|^2 + |\phi^{}_{0}|^2\nonumber\\
& & -\Big( \phi^{*}_{23}\phi^{}_{2}S_{31,1}S_{12,0} +\phi^{*}_{23}\phi^{}_{3}S_{31,0}S_{12,1} +\phi^{*}_{23}\phi^{}_{0} S_{31,3}S_{12,2} + \mathrm{ c.c. }\Big) \nonumber\\
& & +\Big( \phi^{*}_{2}\phi^{}_{3} S_{1,0}S_{0,1} + \phi^{*}_{2}\phi^{}_{0} S_{1,3}S_{0,2} + \phi^{*}_{0}\phi^{}_{3} S_{3,0}S_{2,1} + \mathrm{ c.c. } \Big)
\end{eqnarray}

For the wavefunctions we use the coordinate system defined in Figure~\ref{fig3}, and use $g_i$ to represent the real two-dimensional Gaussian which is the modulus of $\phi_i(\mathbf{r}_1)$, so that the wavefunctions may be written:

\begin{eqnarray}
\phi^{*}_{0} = g_{0} \nonumber\\
\phi^{*}_{2} = g_{2}\mathrm{exp}(id(\surd{3}y - x)) \nonumber\\
\phi^{*}_{3} = g_{3}\mathrm{exp}(id(\surd{3}y + x)) \nonumber\\
\phi^{*}_{23} = g_{23}\mathrm{exp}(id(2\surd{3}y)).
\end{eqnarray}

These expressions reflect the fact that the deflected particles have a longer path length, so have a phase lag with respect to the undeflected beam, and this lag increases with distance away from $x=y=0$.

In calculating an overlap integral, we need to include  $\theta$, the phase of $\phi^{*} \times \phi$,  which varies with position as $2 (s / \sigma) d$ (equation~(\ref{F4}) in the main text) for $S_{0,2}$, $S_{0,3}$, $S_{2,23}$, $S_{2,3}$ \& $S_{3,23}$. For $S_{0,23}$ this phase is $ 2\surd(3) (s / \sigma) d$. In what follows, we express distances in units of $\sigma$ for simplicity of calculation, so that $\phi_0 \propto \mathrm{exp}(-(x^2+y^2)/4)$. Using the fact that $\int \mathrm{exp}(-x^2/2 +Jx) dx = \mathrm{exp}(J^2/2)\int \mathrm{exp}(-x^2/2) dx$, we calculate the overlaps. Those cross terms with negative signs also have an overall positive phase of $ 4 (r/\sigma) \times d = 4kd$, arising from the interaction potential contribution $\theta_i$ to the phase of the 3-particle wavefunctions. This term also corresponds to a phase lag, so $\theta_i$ has the same sign as $\theta$. However, the phase due to the interaction potential cancels for the other cross terms and for the non-cross terms. Hence we obtain the following expressions for the cross terms:

\begin{eqnarray}
-\Big( g_{23}\mathrm{exp}(i2\surd{3}dy) g_{2} \mathrm{exp}(-i\surd{3}dy)\mathrm{exp}(idx)\times \mathrm{exp}(-d^2/8 -2d^2)\mathrm{exp}(2id^2) \times \mathrm{exp}(-3d^2/8 -6d^2)\mathrm{exp}(3id^2)\mathrm{exp}(4ikd) + \mathrm{ c.c. } \Big) \nonumber\\
-\Big( g_{23}\mathrm{exp}(i2\surd{3}dy) g_{3} \mathrm{exp}(-i\surd{3}dy)\mathrm{exp}(-idx)\times \mathrm{exp}(-3d^2/8 -6d^2)\mathrm{exp}(3id^2) \times \mathrm{exp}(-d^2/8 -2d^2)\mathrm{exp}(2id^2)\mathrm{exp}(4ikd) + \mathrm{ c.c. } \Big) \nonumber\\
-\Big( g_{23}\mathrm{exp}(i2\surd{3}dy) g_{0} \times \mathrm{exp}(-d^2/8 -2d^2)\mathrm{exp}(2id^2) \times \mathrm{exp}(-d^2/8 -2d^2)\mathrm{exp}(2id^2)\mathrm{exp}(4ikd) + \mathrm{ c.c. } \Big) \nonumber\\
+\Big( g_{2}\mathrm{exp}(i\surd{3}dy)\mathrm{exp}(-idx) g_{3} \mathrm{exp}(-i\surd{3}dy)\mathrm{exp}(-idx)\times \mathrm{exp}(-d^2/8 -2d^2)\mathrm{exp}(id^2) \times \mathrm{exp}(-d^2/8 -2d^2)\mathrm{exp}(-id^2) + \mathrm{ c.c. } \Big) \nonumber\\
+\Big( g_{2}\mathrm{exp}(i\surd{3}dy)\mathrm{exp}(-idx) g_{0}\times \mathrm{exp}(-d^2/8 -2d^2) \times \mathrm{exp}(-d^2/8 -2d^2)\mathrm{exp}(-id^2) + \mathrm{ c.c. } \Big) \nonumber\\
+\Big( g_{0} g_{3}\mathrm{exp}(-i\surd{3}dy)\mathrm{exp}(-idx)\times \mathrm{exp}(-d^2/8 -2d^2)\mathrm{exp}(id^2) \times \mathrm{exp}(-d^2/8 -2d^2) + \mathrm{ c.c. } \Big)
\end{eqnarray}

Giving finally for the x-y probability distribution for particle 1:

\begin{eqnarray}
P(\mathbf{r}_1) = g_{23}^2 + g_{2}^2 + g_{3}^2 + g_{0}^2\nonumber\\
-2g_{23}g_{2} \mathrm{cos}(\surd{3}dy+dx+5d^2+4kd)\times \mathrm{exp}(-d^2/2 -8d^2) \nonumber\\
-2g_{23}g_{3} \mathrm{cos}(\surd{3}dy-dx+5d^2+4kd)\times \mathrm{exp}(-d^2/2 -8d^2) \nonumber\\
-2 g_{23}g_{0} \mathrm{cos}(2\surd{3}dy+4d^2+4kd)\times \mathrm{exp}(-d^2/4 -4d^2) \nonumber\\
+2g_{2}g_{3}\mathrm{cos}(2dx)\times \mathrm{exp}(-d^2/4 -4d^2) \nonumber\\
+2g_{2}g_{0}\mathrm{cos}(\surd{3}dy-dx-d^2)\times \mathrm{exp}(-d^2/4 -4d^2) \nonumber\\
+2g_{0}g_{3}\mathrm{cos}(-\surd{3}dy-dx+d^2)\times \mathrm{exp}(-d^2/4 -4d^2)
\end{eqnarray}

The equivalent expression ignoring phase differences is:
\begin{eqnarray}
P(\mathbf{r}_1) = g_{23}^2 + g_{2}^2 + g_{3}^2 + g_{0}^2\nonumber\\
-2g_{23}g_{2} \times \mathrm{exp}(-d^2/2) \nonumber\\
-2g_{23}g_{3} \times \mathrm{exp}(-d^2/2) \nonumber\\
-2 g_{23}g_{0} \times \mathrm{exp}(-d^2/4) \nonumber\\
+2g_{2}g_{3} \times \mathrm{exp}(-d^2/4) \nonumber\\
+2g_{2}g_{0} \times \mathrm{exp}(-d^2/4) \nonumber\\
+2g_{0}g_{3} \times \mathrm{exp}(-d^2/4)
\end{eqnarray}

In this calculation, we have assumed that the phase differences introduced by the interactions have zero variance. However, $\theta_i$ depends on the distance $r$ between particles and this has a standard deviation $\surd2 \sigma$. Hence, in an ensemble of measurements, we would expect a spread of values of phase difference $\sigma_{\theta} \propto \sigma / r \times \theta_i$  in all the cross terms above. Averaging over this will multiply each cross term by $\mathrm{exp}(-(\sigma_{\theta})^2/2)$. This, or experimentally-produced phase errors, will tend to remove the effect of the cross terms, particularly at large $d$.

\subsection*{Is a quantum PHE experiment feasible with electrons?}

Rather than writing the force of interaction between two particles in the general form  $2 \Delta V / r$, we may use the actual Coulomb force between two electrons: $F = e^2/4 \pi \epsilon_0 r^2$. Inserting this into equation~(\ref{F8}), with a little manipulation we obtain:

\begin{equation}
 \Delta r = \frac{1}{2}   \frac{me^2}{4 \pi \epsilon_0 \hbar^2} (\ell \lambda)^2 / 4 \pi^2   r^2. \label{T1}
\end{equation}

We notice a combination of constants which is the Bohr radius $a_0$, and using equation~(\ref{F1}), this becomes:

\begin{equation}
 \Delta r = \frac{1}{2}   \frac{\sigma^4 }{a_0 r^2}. \label{T2}
\end{equation}
We may further express this in terms of the interaction strength $d$:

\begin{equation}
 d =  \frac{\sigma^2 }{ r^2} \frac{\sigma }{2a_0}, \label{T3}
\end{equation}

\noindent where $2a_0$ has the numerical value $1.06 \times 10^{-10}$ m.  Now, in order that the particles should be distinguishable, we need  $ r/\sigma >>1$ - at least $\sim 5$. Also, $d \times r/\sigma$ has to be small to avoid large phase differences, which remove the quantum PHE. Figure~\ref{fig6} in the main text indicates that for $ r/\sigma=5$, $d$ should not be larger than  $\sim 0.005$.  Inserting these numbers into equation~(\ref{T3}) gives:

\begin{equation}
 \sigma \sim 0.25a_0 \sim 3 \times 10^{-11} \mathrm{m}. \label{T4}
\end{equation}

We also need  $\sigma << \ell$. A pair of values satisfying this and Eqn.~(\ref{T4}) are $\lambda \sim 10^{-12} \mathrm{m}$ (this corresponds to electrons with kinetic energy $\sim 40$ keV) and $\ell \sim 5 \times 10^{-9} \mathrm{m}$. The values of $ \sigma \sim 3 \times 10^{-11} \mathrm{m}$ and $d \sim 0.005$ give $\Delta r \sim 2 \times 10^{-13} \mathrm{m}$. These values (and any others satisfying the the restrictions above) indicate that it would be extremely challenging to obtain the required measurement accuracy for $\Delta r$ and also to control the phases of the electrons. Another challenge would be to design an experiment where the interaction strength could be accurately controlled and tuned precisely to a required value  Therefore, we believe that performing this experiment with electrons would probably be impossible and it follows from the above calculation that it would be even more of a challenge if charged particles of greater mass were employed.

A different quantum PHE thought-experiment involving atoms is outlined in Ref.~[1] in the main text. It is proposed that interactions between the atoms will be revealed via a small frequency change of photons emitted by them. The energy of atomic interaction will certainly exceed the energy change of the emitted photons, and this interaction will entail both deflections and phase shifts in the atom wavefunctions. Without an explicit case to consider, further analysis is difficult, but we expect that an atom experiment would prove equally challenging as an electron experiment.

 \end{document}